\documentclass{PoS}

\newcommand\ignore[1]{}

\newcommand\be{\begin{equation}}
\newcommand\ee{\end{equation}}
\newcommand\bea{\begin{eqnarray}}
\newcommand\eea{\end{eqnarray}}

\renewcommand{\>}{\rangle}

\title{Improved Lattice Radial Quantization}

\ShortTitle{FEM  $\phi^4$ }

\author{\speaker{Richard C. Brower}\thanks{RCB. acknowledges support under DOE grants DE-FG02-91ER40676,
DE-FC02-06ER41440, and NSF grants OCI-0749317, OCI-0749202.  GTF acknowledges partial support by
the NSF under grant NSF PHY-1100905. }\\
      Department of Physics\\
       Boston University, 590 Commonwealth Ave, Boston, MA 02215, USA\\
        E-mail: \email{brower@bu.edu}}

\author{Michael Cheng\\
 Center for Computational Science \\
       Boston University, 3 Cummington Mall, Boston , MA 02215, USA\\
        E-mail: \email{micheng@bu.edu}}

\author{George T. Fleming\\
       Department of Phsics\\
     Yale University, Sloane Laboratory, New Haven, CT 60520, USA \\
        E-mail: \email{George.fleming@yale.edu}}

\abstract{Lattice radial quantization was proposed in a
recent paper by Brower, Fleming and Neuberger~\cite{Brower:2012vg} as  a  nonperturbative
method especially suited to numerically solve Euclidean  conformal
field theories.  The lessons learned from the
lattice radial quantization of the 3D Ising model
on a longitudinal cylinder with  2D Icosahedral cross-section suggested the need
for an improved discretization.   We consider here
the use of the Finite Element Methods(FEM)  to descretize
the universally-equivalent  $\phi^4$ Lagrangian on $\mathbb R \times
\mathbb S^2$.  It is argued that this lattice
regularization will approach  the exact conformal theory at the Wilson-Fisher fixed
point in the continuum. Numerical tests are underway to support this
conjecture. }

\FullConference{31st International Symposium on Lattice Field Theory - LATTICE 2013\\
		July 29 - August 3, 2013\\
		Mainz, Germany}

\begin{document}


\section{Introduction}
Conformal or near conformal behavior in field theory lies at the heart of many challenging theoretical
and phenomenological problems. Models for possible strong dynamics for electro-weak symmetry breaking
as a replacement of the elementary Higgs of the Standard Model are  often built on near-conformal
theories. A large variety of extra-dimensional models use  the AdS/CFT correspondence to
introduce large scale separations. However conventional lattice methods near
an IR conformal fixed point are difficult, precisely because of the growing
separation of the length scales between the UV and IR.  A recent paper by  Brower, Fleming and Neuberger(BFN) 
explored replacing the traditional Euclidean lattice in favor of one
suited to {\em Radial Quantization} ~\cite{Brower:2012vg}. In radial quantization the dilatation operator plays the role of the
Hamiltonian. The
potential advantage for lattice simulations is now the  dilatation
operator generates translations in $\log r$ so a finite radial lattice
separates scales {\em exponentially} in
the number of lattice sites. 

For an exactly conformal field theory, the  idea is straightforward.   The flat metric  for
any Euclidean field theory on $\mathbb R^D$ can obviously be expressed in radial coordinates,
\begin{equation}
 ds^2 = g_{\mu\nu}dx^\mu dx ^\nu =  r^2_0 e^{ 2t} ( dt^2 + d\Omega^2_{D-1} ) \; , 
\end{equation}
where $t = \log(r/r_0)$, introducing an arbitrary reference scale $r_0$, and
where $d\Omega^2_{D-1}$ is the metric on the $\mathbb S^{D-1}$ sphere of unit
radius.  However for an exactly conformal
field theory, 
the Weyl
transformation  can also be 
used to  remove
the conformal factor, $e^{2t}$.
As a result, a CFT can be legitimately mapped from the  D-dimensional  Euclidean space $\mathbb
R^D$ to perform radial quantization on a cylindrical manifold, $\mathbb R\times \mathbb S^{D-1}$.

On $\mathbb R\times \mathbb S^{D-1}$,  dilatations, inversion at the origin, and
rotations on  $\mathbb  S^{D-1}$ are manifest. The extension
to the full conformal group O(D+1,1), including Poincar\'e invariance, is now a consequence of the dynamics.
Ref.~\cite{Brower:2012vg}  explores the circumstances when, and when not, full
conformal invariance is achieved. As a first example,
the large $N$ solution to the 2D O($N$) sigma
model, which (like massless QCD)  has a  classically conformal
Lagrangian with a quantum conformal anomaly, is considered.
Comparing the traditional Euclidean $\mathbb R^2$
quantization to  radial quantization on $\mathbb R\times
\mathbb S^{1}$, it was recognized  that the quantum anomaly forces the traditional
 $\mathbb R^2$ quantization to choose the Lorentz symmetry (O(2) rotations),
whereas   radial $\mathbb R\times \mathbb S^{1}$  quantization  chooses
 dilatation invariance. Neither quantization scheme had the full
 conformal symmetry.  Repeating this exercise for the 3D O($N$)
 model, having no anomaly, apparently gives the full conformal
 symmetry group, O(4,1), for both.  Calculations on the large $N$
 expansion in 3D are underway to prove this.

A second example considered in ref.~\cite{Brower:2012vg} was the lattice implementation of 
radial quantization  for the 3D Ising model at the Wilson-Fisher fixed
point. A discrete lattice was used, on a cylinder with longitudinal coordinate
in $t = \log r$ and  uniform equilateral triangular refinement of the icosahedral approximation to
$\mathbb S^2$. Here the numerical methods proved generally quite accurate, but
as noted in  the conclusion, there were small
departures away from the full restoration of the conformal group  in the
continuum. In particular, the scaling dimensions for the two irreducible icosahedral representations 
for the 3rd descendant of the $Z_2$ odd primary did not converge to a
single $l = 3$ representation of O(3) in
the continuum. Apparently this lattice implementation in the
continuum limit yields a  radial quantization of a critical Ising model on a transverse  icosahedron, which quite naturally  failed
to realize the full conformal symmetry. In this presentation, we seek to
remove this obstruction  by the use of a Finite Element
Method (FEM)  for the  universally-equivalent $\phi^4$
Lagrangian in order to allow convergence to O(3) symmetry on
$\mathbb S^2$. This improved radial quantization,  we conjecture, will approach the
exact conformal field theory at the Wilson-Fisher fixed point with
full O(4,1) conformal symmetry in the
continuum limit without fine-tuning of irrelevant operators on the lattice.

\section{Finite Element Lagrangian}

The  problem we face is how to put a quantum field theory on
 a tessellation of a curved manifold. The finite element method (FEM) is an established mathematical method
to discretize differential equations on an irregular mesh~\cite{Strang}. 
It is not clear
that a variation of FEM can be formulated for  the Ising model,because there is no notion of a locally smooth field
configuration. Therefore, we  first replace our Ising model
by the universally-equivalent $\phi^4$ Lagrangian.  The Lagrangian 
on a  smooth manifold is given by 
\be
S = \int d^Dx \sqrt{-g} [\frac{1}{2}g^{\mu \nu} \partial_\mu \phi \partial_\nu
\phi  + \lambda( \phi^2 - \frac{\mu^2}{2 \lambda}) ^2] \; .
\ee
On a flat  $\mathbb R^3$ manifold, a suitable discretization by  a
regular  cubic lattice is achieved by replacing the kinetic term by a
nearest neighbor finite difference
approximation. The justification for this can be understood
by performing a local Taylor series expansion at each site $x$ for the
kinetic term in the classical equation of motion, 
\be
a^{-2}\sum_{\pm \hat\mu} [\phi(x) - \phi(x + a \hat \mu)] \simeq
-\nabla^2 \phi(x) + O(a^2)  \; ,
\label{eq:RDtaylor}
\ee
and observing that rotational symmetry is restored up to $O(a^2)$
corrections and that at the quantum level this introduces symmetry breaking only
by irrelevant operators.  The reason for this success is traced to the
fact that the hypercubic group preserves an infinitely growing
subgroup of the Poincar\'e group (cubic rotations plus discrete
translations).  Poincar\'e invariance is restored as an accidental
symmetry.

\begin{figure}[ht]
\vskip -0.5cm
\centering
\includegraphics[width=0.6\textwidth]{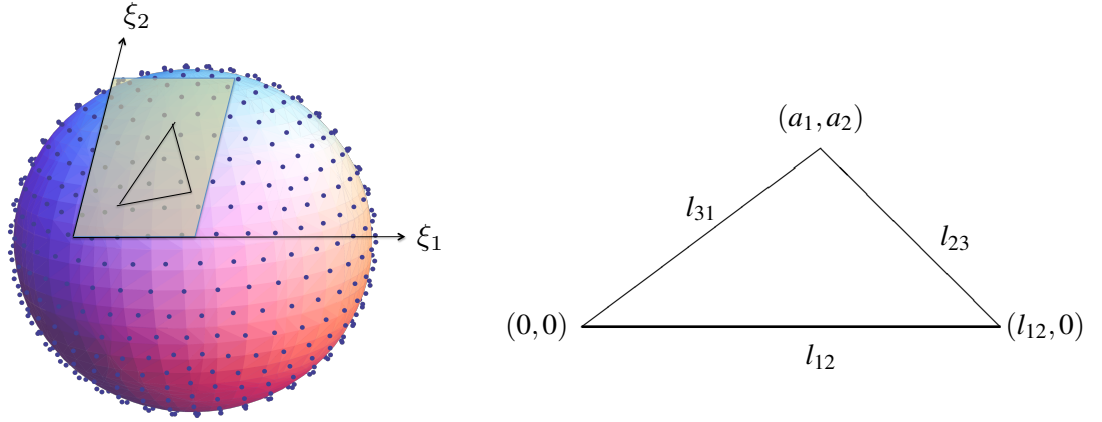}
\setlength{\unitlength}{.31in}
\begin{picture}(7,1)(0,-2.5)
\linethickness{1pt}
\put(0,0){\line(1,0){7}}
\put(7,0){\line(-1,1){3}}
\put(0,0){\line(4,3){4}}
\put(-0.75,0){\makebox(0,0){$(0,0)$}}
\put(7.75,0){\makebox(0,0){$(l_{12},0)$}}
\put(4,3.5){\makebox(0,0){$(a_1,a_2)$}}
\put(4,-.5){\makebox(0,0){$l_{12}$}}
\put(6.25,1.5){\makebox(0,0){$l_{23}$}}
\put(2,2){\makebox(0,0){$l_{31}$}}
\end{picture}
\vskip -0.5cm
\caption{\label{fig:FEM}. The coordinates on the  sphere $x^\mu$ for
  each spherical triangle are mapped one to one by radial projection on to the
tangent plane Cartesian coordinates $\xi_a(x)$. }
\end{figure}

On a sphere there is no analogous infinite subgroup of the rotation group, so
a new method is needed.  Instead we consider a finite element
approach to the lattice discretization of the  $\mathbb R
\times S^2$ manifold to address this problem.  The simplest
option is to use first order piecewise linear  finite elements on 
each triangle projected locally onto its tangent plane as illustrated in Fig.~\ref{fig:FEM}. Introducing local Cartesian
coordinates, $\xi_a$, on each tangent plane,  we  approximate
the field $\phi(x)$ as a sum over the three non-zero linear
elements $\phi(x) = \sum_i \phi_i \; L_i(\xi_1,\xi_2) $ on each
triangle, where the projection is defined by the function $x^\mu(\xi_a)$.  For example if we label
the vertices by  $(1,2,3)$   we may orient the coordinate system with
vertices at $(\xi_1,\xi_2)= (0,0), (l_{12},0)$ and $(a_1,a_2)$
respectively by an appropriate 2D Euclidean  transformation (rotation
plus translation). Now the
elements take the explicit form,
$$
L_1(\xi_a) = [l_{12}-\xi_1  - (l_{12}-\xi_1) (\xi_2/a_2)]/l_{12}\; ,\; \;  \;  L_2(\xi_a)
=[ \xi_1 - a_1(y/a_2)]/l_{12} \;  , \;  \;  \;  L_3(\xi_a) =
\xi_2/a_2 
$$
such that $L_i(x,y)$  is 1 on vertex i and zero on the other two. 
A straightforward calculation gives the  contribution of this
triangle to the kinetic term,
\be
\int_{A}  d^2\xi \; \partial_a  \phi(\xi)  \partial_a \phi(\xi)
= \frac{1}{8 A_{123}}[(l^2_{23} + l^2_{31} - l^2_{12})(\phi_1 - \phi_2)^2 +
\mbox{cyclic} ] \; ,
\label{eq:FEMtriangle}
\ee
 where $l_{ij}$  are the lengths of the edges  and
$A_{ijk}$ is  the area of the $(1,2,3)$  spherical triangle projected
onto the tangent plane.  Note that the resulting FEM integral is indeed invariant
under  Euclidean group (rotation and translations) on the 2D tangent
plane, as assumed in our choice of tangent plane coordinates.
To complete the FEM expression, the lattice for the radial coordinate was chosen 
as before with a uniform spacing and the kinetic term, $(\partial_t
\phi)^2$, represented by a  finite difference. To account for the
density $\sqrt{-g}$ in the continuum both the radial kinetic term
and the  potential terms are weighted by the local volume element, $\omega_x$,
computed as   $1/3$ the sum of the  areas of triangles adjacent to the site
$x$.  The resulting  discrete FEM  action is given by
\be
S = \frac{1}{2}\sum_{t,\<x \; y\>} K_{x,y}(\phi_{t,x} - \phi_{t,y})^2  +
\frac{1}{2} \sum_{t,x} \omega_x  
(\phi_{t+1,x} - \phi_{t,x})^2 + \lambda\sum_{t,x} \omega_x  
(\phi_{t,x}^2 - \mu^2/2 \lambda)^2 \; ,
\ee
where $(t,x)$  denotes lattice sites:
$t = 0,\cdots,  N_t-1$ enumerates the spheres and $x = 1, \cdots,
2 + 10 s^2$, the sites on each sphere with refinement level $s$,
starting with   $s =
1$  for  the unrefined icosahedron. Consequently $1/s$ plays  the role of the
lattice spacing: $a \sim 1/s$.  $K_{x,y}$ are the FEM weights for each
link $\<x\; y\>$ on the
triangulated ${\mathbb  S}^2$ computed from the adjacent
triangles using Eq.~\ref {eq:FEMtriangle}. 
One can prove that the Legendre functions, $Y(\theta_x,\phi_x) \equiv
Y_{lm}(\hat x)$,  evaluated on the
vertices on the sphere, converge to the continuum orthonormal basis,
\be
\lim_{s \rightarrow \infty} \sum^{2 + 10 s^2}_{x = 1} \omega_x
Y^*(\hat x)_{l'm'} Y_{lm} (\hat x) = 4 \pi \delta_{l'l} \delta_{m'm} 
\ee
in the zero lattice spacing limit.

\section{Spectral Properties of the Laplacian on FEM Sphere}

Several comments are in order. Contrary
to the finite difference approximation on a  hypercubic lattice 
in ${\mathbb R}^D$ in Eq.~\ref{eq:RDtaylor}  above, the Taylor series expansion
of the kinetic term on FEM sphere does {\em not} yield
a local rotationally invariant Laplace operator $\nabla^2 \phi +
O(a^2)$.
 It is straightforward to perform an explicit Taylor series expansion,
\be
a^{-2}\sum_{y \in \<x y\>}K(x,y)[\phi(x) - \phi(y)] \simeq 
- c_{\mu \nu} \partial_\mu \phi(x) \partial_\nu  \phi(x) + O(a^2) 
\ee
to  evaluate the symmetric matrix, $c_{\mu \nu} $, as a function of
the  length of sides of the adjacent triangles. Rotational invariance
is generally  violated at leading order ($c_{11} \ne c_{22}, c_{12} \ne 0$),
except for a few vertices  lying on  the symmetry axes of  the original
icosahedron. However, local
rotational invariance of this Taylor expansion is not essential to
a faithful continuum limit for lattice field
theory.    For a   sequence of
more and more refined lattice operators, it is sufficient to require  that  the  spectrum  of
eigenvalues, well below the cut-off,  converge to the exact
continuum values.   Of course on the regular hypercubic lattice,
this spectral condition is also  trivially satisfied as can be see by
diagonalizing Eq.~\ref {eq:RDtaylor} in Fourier space: $\sum_\mu
4 a^{-2} \sin^2(a k_\mu/2) \simeq k^2 + O(a^2)$.
What we need is  the analogous spectral property for our FEM
discrete Laplacian on the sphere.

\begin{figure}[ht]
\centering
\includegraphics[width=0.45\textwidth]{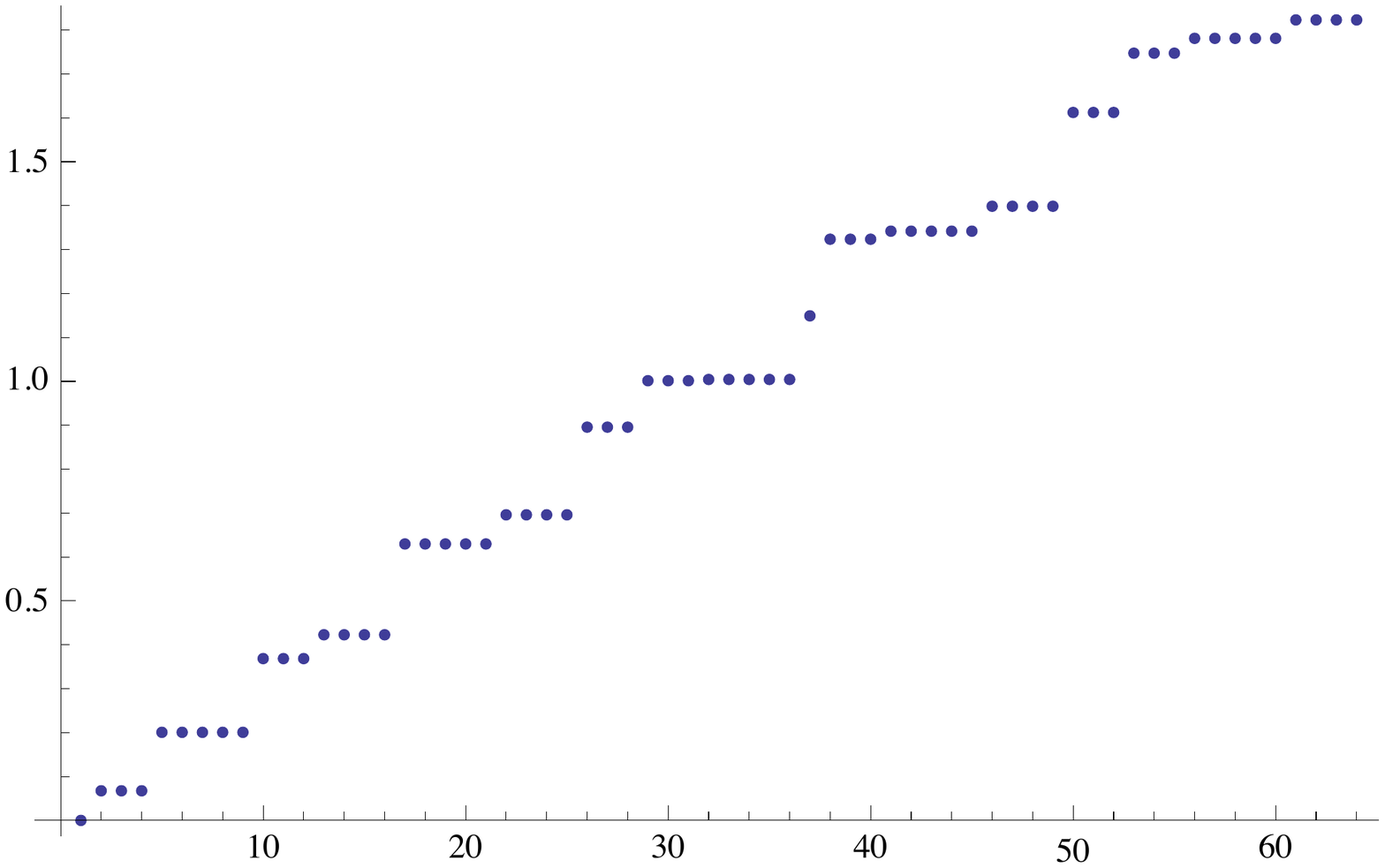}
\hskip 1cm
\includegraphics[width=0.45\textwidth]{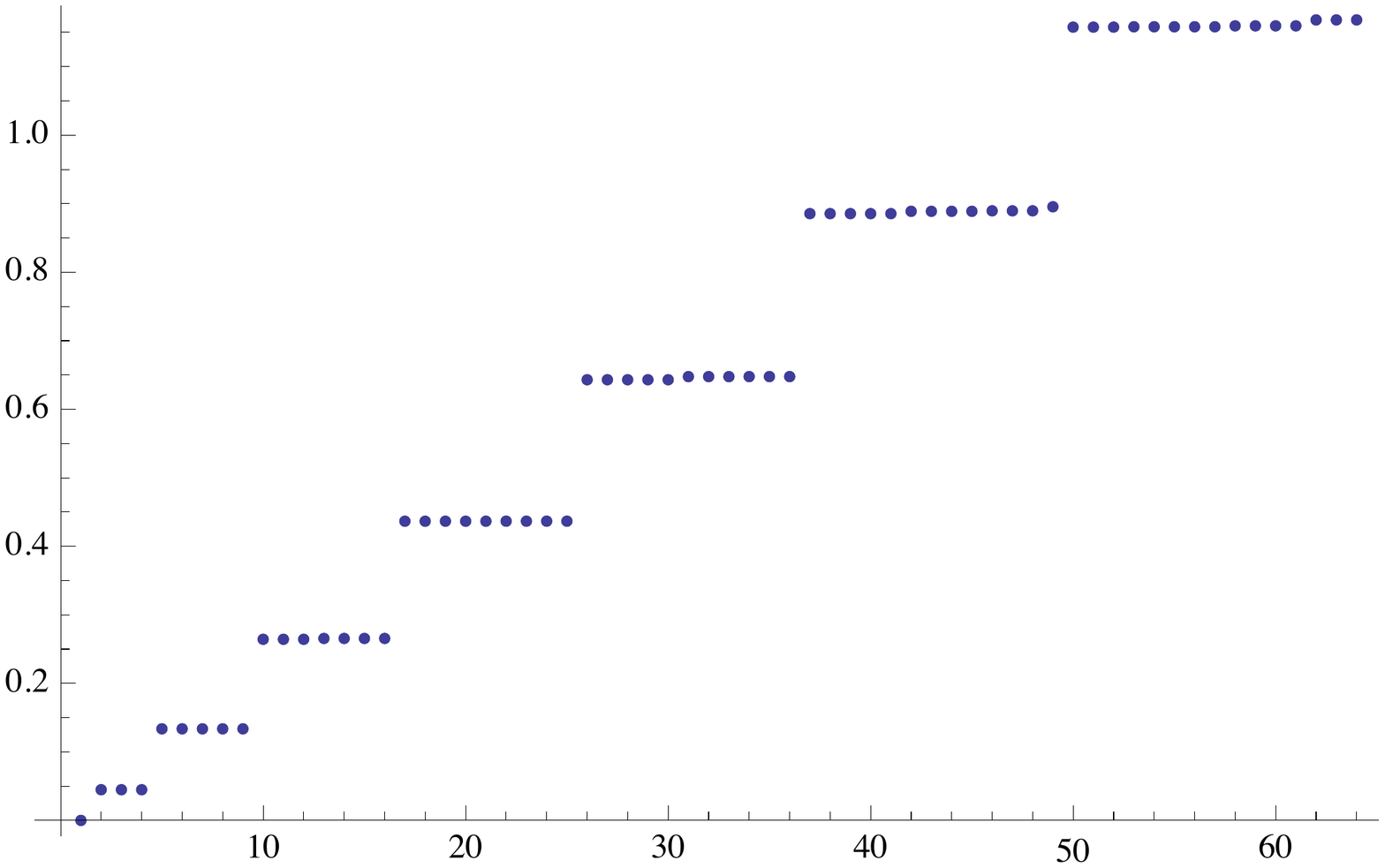}
\caption{\label{fig:UnifromFEM}. For $s = 8$, the comparison  of 
  the first 64 eigenvalues for  the unimproved spectrum
  discrete  icosahedral Laplace operator on the left vs the FEM operator 
on the sphere on the right. }
\end{figure}

Indeed in  FEM literature, there are many theorems on
 spectral convergence. The general take-away lesson is that  if the sequence of grid refinements
have simplices that uniformly shrink to zero (\textit{e.g.}\ a 2D area 
bounded by $O(a^2)$)  and obey an appropriate ``shape-regular''
condition (\textit{e.g.}\ in 2D bounds on ratios of angles) then the
spectra of the FEM  Laplace operator will
converge as $a^{2p}$ for $p$-order finite elements  and their
eigenvectors  will converge to their continuum form
as $a^p$.  To test this FEM lore for our extension to  the sphere, we
are studying the spectral properties numerically. The 
generalized eigenvalue condition on the FEM Laplacian for a sphere is: $K_{xy}\phi_n(y) =
\lambda_n \omega_x \phi_n(x)$. In Fig.\ref{fig:UnifromFEM}, we compare the
lowest 64 eigenvalues for the unweighted ($K_{xy} = 1$)  finite
difference operator, used in Ref.~\cite{Brower:2012vg}, with that of the improved FEM operator. The $l = 0$ (Singlet)
, $l= 1$ (Triplet) and $l = 2$ (Quintet) are in irreducible representations
of the icosahedral group and are in excellent agreement for both. However
for the $l \ge 3$ eigenvalues, the $2l + 1$ degeneracy is clearly much improved by the finite element method even
with a  modest refinement level of  $s = 8$.

\begin{figure}[ht]
\centering
\includegraphics[width=0.45\textwidth]{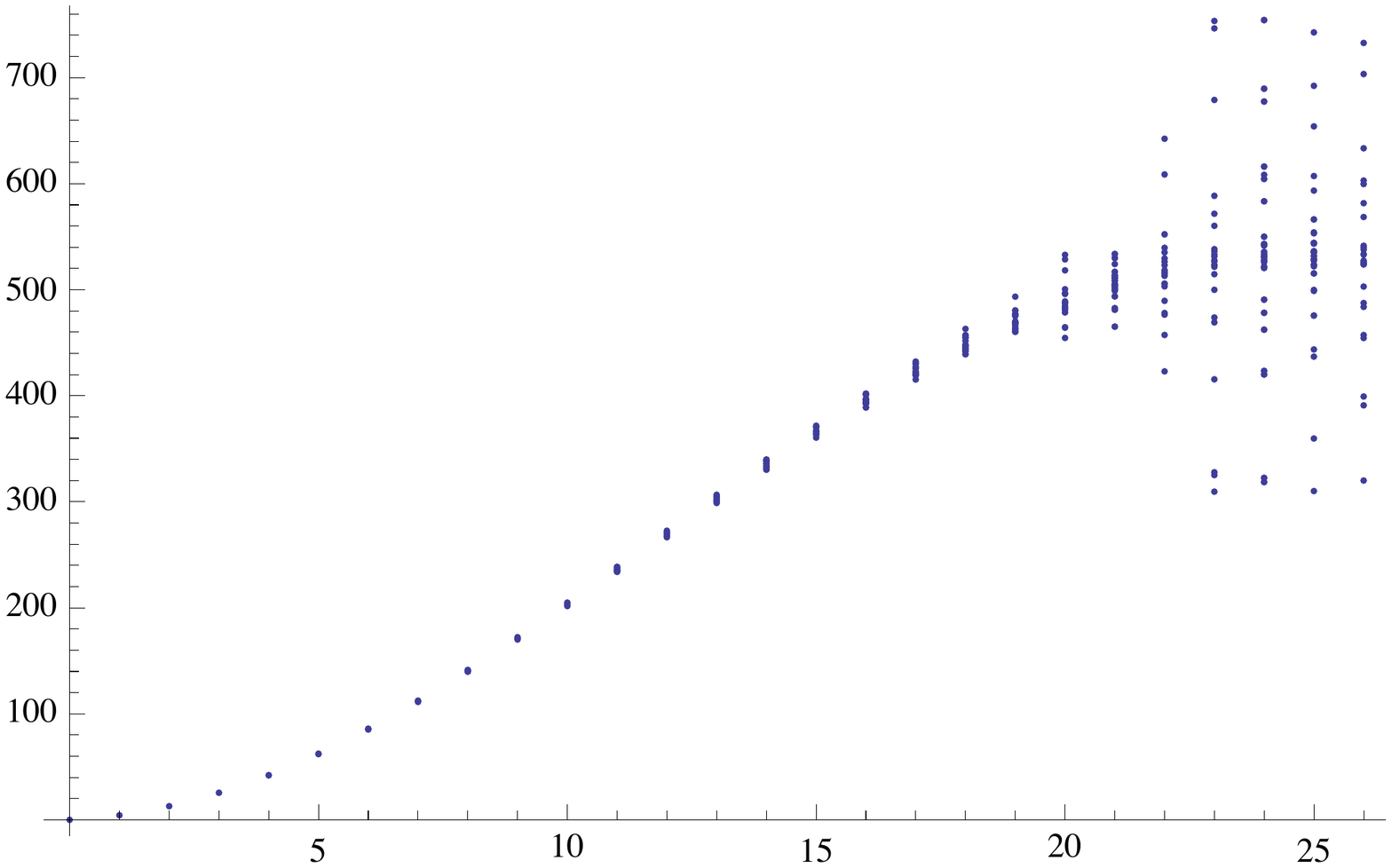}
\hskip 1cm
\includegraphics[width=0.45\textwidth]{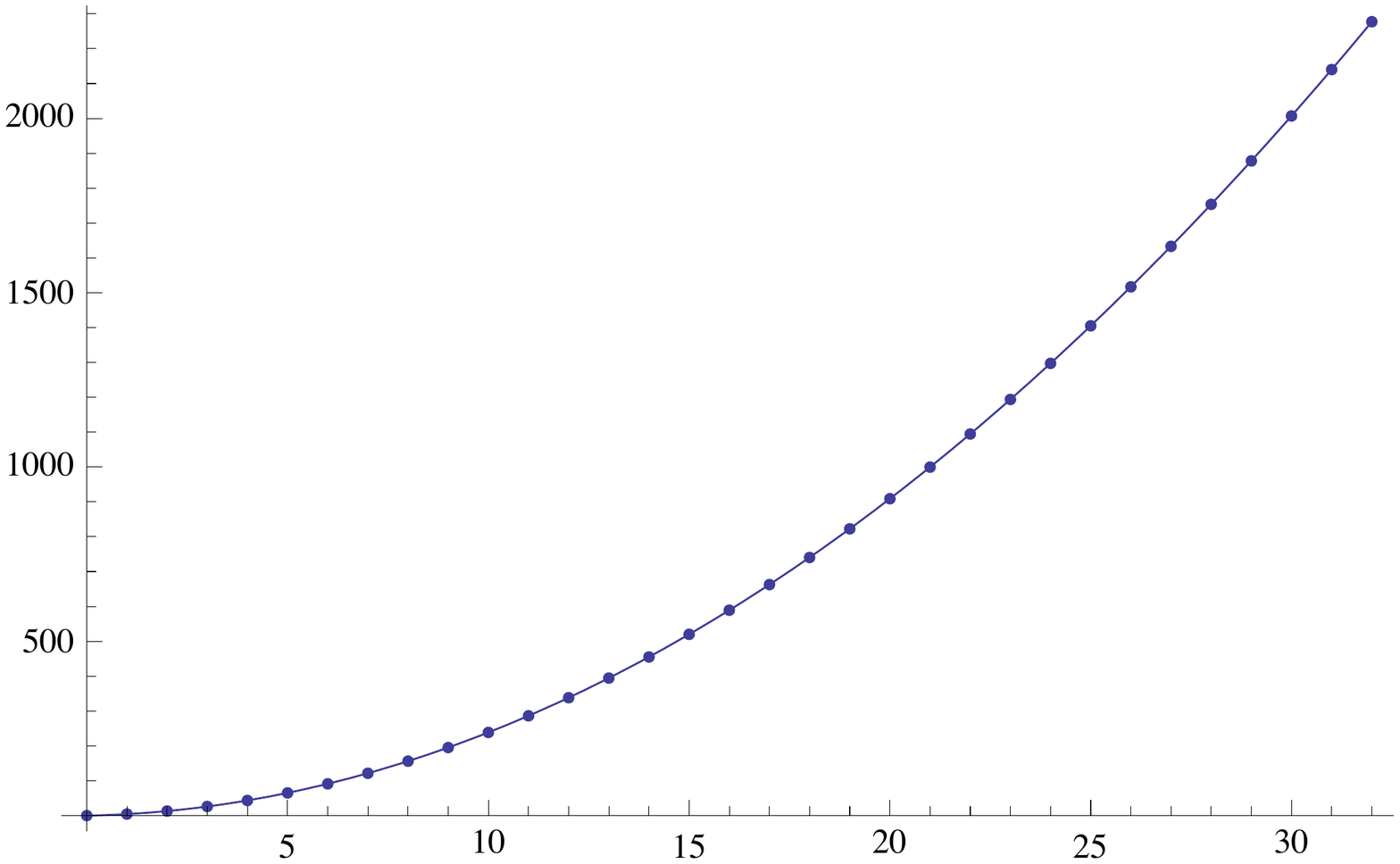}
\caption{\label{fig:FEMdiag}. On the left  all the diagonal elements, 
  $c_{lm} = \sum_{\<x\;y\>}Y^*_{lm}(\hat x) K(x,y)  Y_{lm}(\hat y)$, are plotted  against $l$ for $s
  = 8$ and on the right $k_{lm}$ averaged over $m$ 
 are fitted to $l + 1.00012 \; l^2 - 1.34281\times 10^{-7} \; l^3 - 0.57244
\times 10^{-7} \; l^4$   for $s =  128$   and $l \le 32$.}
\end{figure}

In Fig.~\ref{fig:FEMdiag}, we plot the diagonal matrix elements,
$c_{lm} = \sum_{\<x\;y\>}Y^*_{lm}(\hat x) K(x,y)  Y_{lm}(\hat y)$, as a
function of $l$.  On the left, for $s = 8$,  all $2 + 10 s^2 = 642$
diagonal elements are included and on the right for $s = 128$ the diagonal
elements  for $l = 0,\cdots,32$ averaged over $m$ match the continuum form, $l(l+1)$,  up to $O(10^{-4})$ corrections.
 A more thorough analysis of
the rate of convergence will be reported in the future, where we will
show that this lattice discretization of the Lagrangian on $\mathbb R
\times \mathbb S^2$ converges to the continuum for all modes a finite fraction below
the  cut-off.  We conjecture that this is sufficient
to guarantee the correct  continuum universality at the
Wilson-Fisher fixed point. 

In a series of pioneering papers, Christ, Friedberg and
Lee~\cite{CFL1,CFL2} developed  a similar approach to place
quantum field theory on random simplicial lattices in flat space.
We note however that their application to random lattices  violates
the ``shape-regular'' constraint needed here to ensure good spectral
properties. 


\section{Discussion}

It is plausible that  FEM lattice radial quantization
of $\phi^4$ considered here will recover the exact CFT with full O(4,1)
invariance at the Wilson-Fisher fixed point in the continuum limit. 
Still  both analytical and numerical methods should be pursued  to
test this conjecture. We note that even in the continuum, the map from  ${\mathbb R}^3$ to  $\mathbb R\times \mathbb
S^{2}$ raises questions for  $\phi^4$ theory, since 
both $\mu$  and $\lambda$ have dimension of mass.  The classical
Lagrangian on $\mathbb R^3$ is not conformally invariant so  after
performing the Weyl transformation,
 we have simply dropped the radial dependence in  $r \mu = r_0  e^t \mu$ and
$r  \lambda = r_0 e^t \lambda $, replacing them by t-independent   dimensionless parameters.  The implicit
assumption we have made is that both  traditional $\mathbb R^3$  and radial $\mathbb R \times
\mathbb S^2$ quantum theories so defined have identical CFT's at
their  Wilson-
Fisher fixed point in spite of the fact that each have inequivalent  deformations
for  $\mu$ and $\lambda$  away from the fixed point. To test this idea,
we are now  performing analytical calculations   for   the
3D  non-compact O($N$) model for both the traditional and radial
quantization in the large N
expansion.   While this will not prove our conjecture for  $N = 1$, it
is a useful first step. 
\begin{figure}[ht]
\centering
\vskip -0.5 cm
\includegraphics[width=0.35\textwidth]{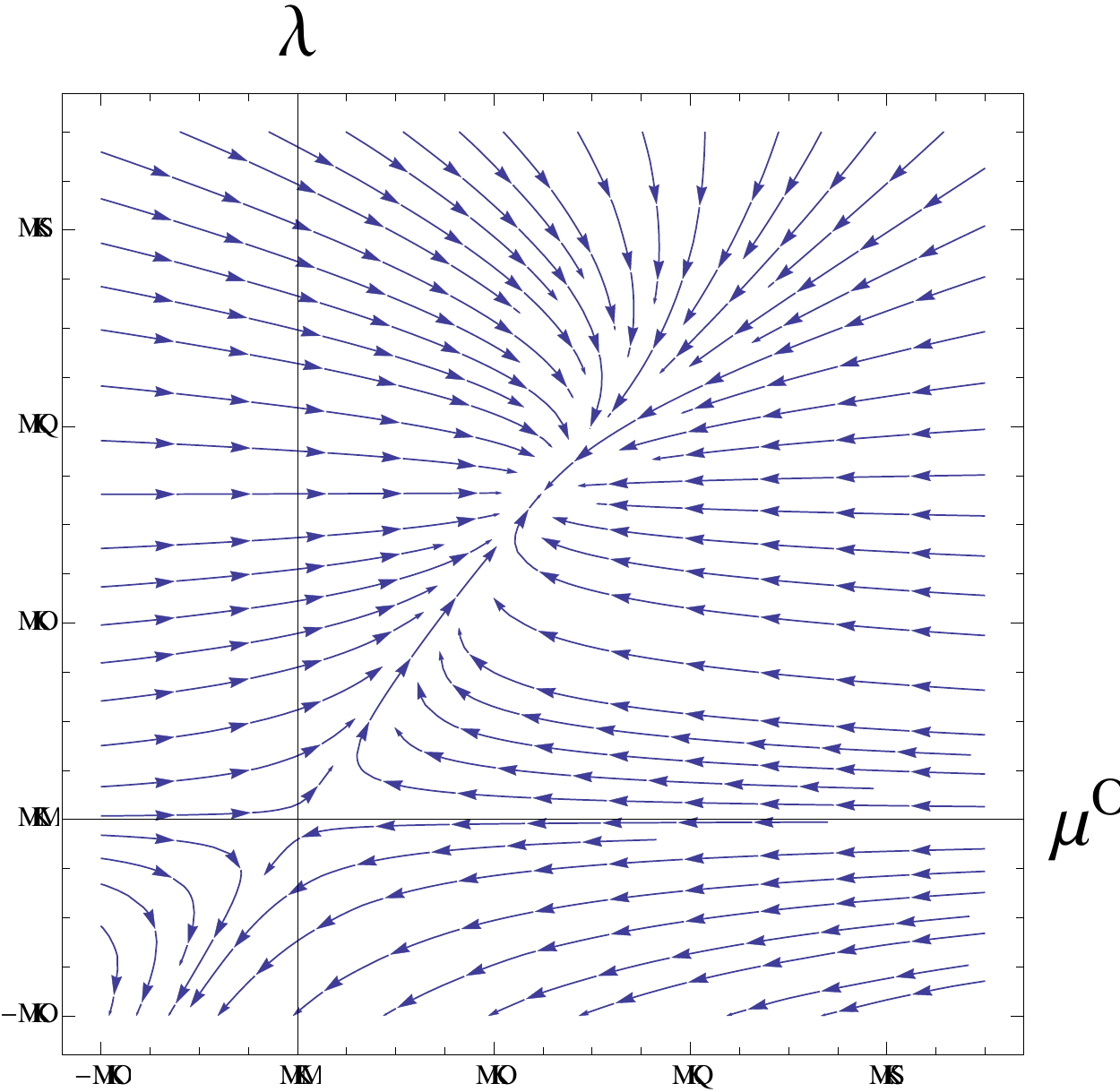}
\raisebox{-0.16\height}{\includegraphics[width=0.53\textwidth]{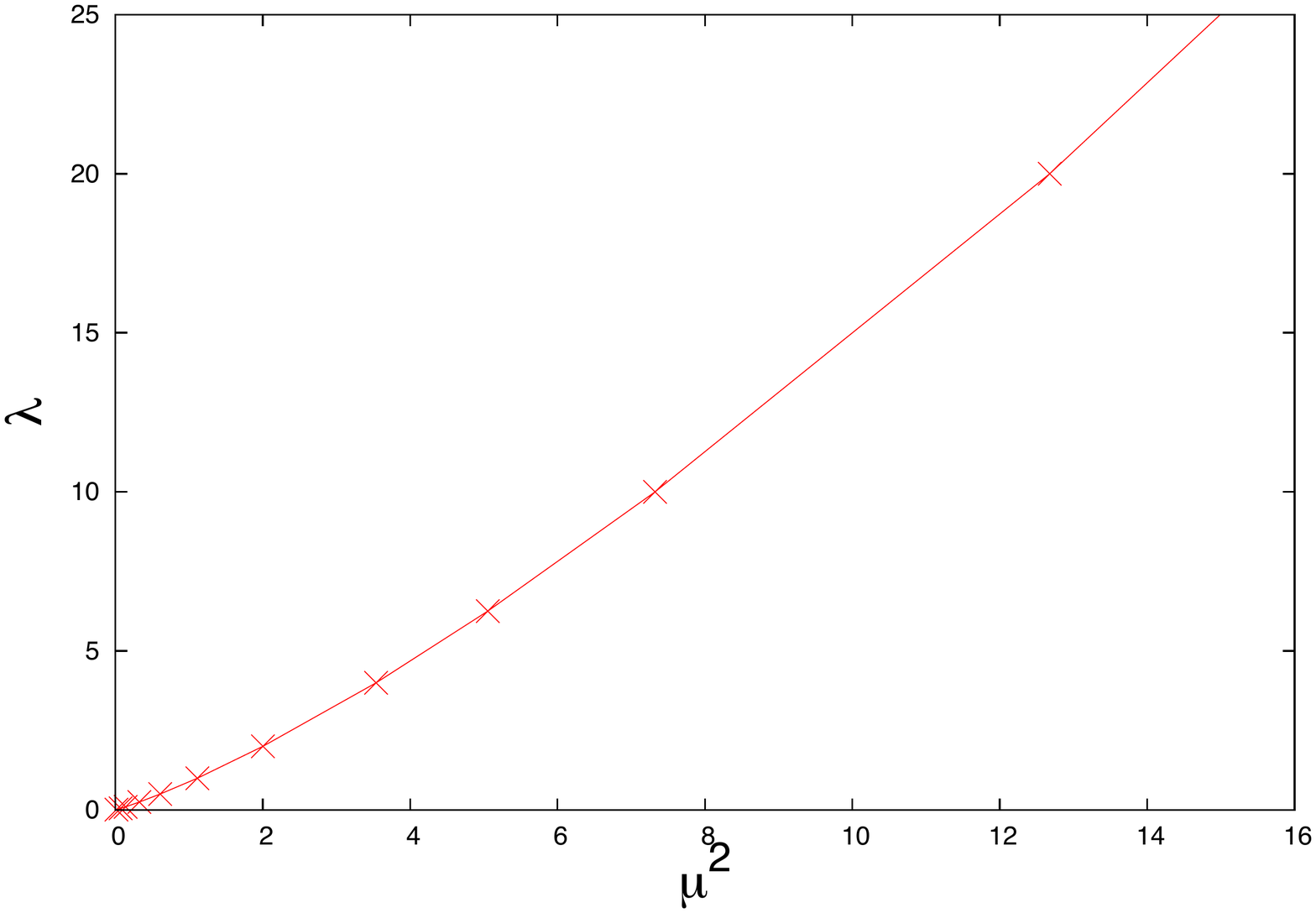}}
\vskip -0.5cm
\caption{\label{fig:CriticalSurface}  On the left is the 
RG flow in the  $(\lambda,\mu^2)$  phase plan  in the continuum epsilon expansion on
$\mathbb R^3$ compared on the right with
the critical surface for the lattice radial theory on $\mathbb R
\times \mathbb S^2$.}
\end{figure}

To further test  our lattice  FEM construction for radial quantization, we are  also pursuing Monte Carlos simulations using the mixed cluster/Metropolis
algorithm of Brower and Tamayo~\cite{Brower:1989mt}.  To date we have located the critical surface
in the bare $(\mu^2,\lambda)$ plane as illustrated in
Fig.~\ref{fig:CriticalSurface}  on the right.   On the left we compare
this with the  phase plane  of $\phi^4$ in the
continuum on $\mathbb R^3$ determined with the $\epsilon$ expansion. The  similarity between the two critical
surfaces is reassuring  but we are just beginning to do high
precision simulations to strength this comparison.  To locate the fixed
point in the radial quantized  critical surface, we are employing the 
methods of Hasenbusch~\cite{Hasenbusch:1999zz, Hasenbusch:1999mw}. 
By setting $\lambda$ to its fix point  value, we seek
an improved action to help to compute more accurately
the anomalous dimensions (or critical exponents) in both the even and
odd $Z_2$ sectors.  Our first goal is to determine if the defect in the 3rd descendant found on the Ising icosahedron is
removed by the FEM improvement action when extrapolated to the  continuum. 
 
We are also generalizing the FEM method to place both gauge fields and fermions
on $\mathbb S^{D-1}$ in order to be able to apply radial quantization to
a wide class of gauge theories with fermionic matter.   For the scalar field
the generalization of FEM  to 4D field theories on  $\mathbb R \times
\mathbb S^3$  using tetrahedral simplices is not difficult.  However
the  generalization of FEM methods to higher spin fermionic and gauge
field is more involved. It requires a more detailed use of geometric objects, most importantly the  verbein $e^a_\mu(x)
= \partial_\mu \xi^a(x) $, relating the curved manifold to the local tangent
plane, familiar to the application of continuum  field theory in
general relativity.  Nonetheless we are optimistic that the FEM can be
applied very generally to lattice radial quantization  and that we
can in due time explore numerically a wide range of consequences for
conformal and IR conformal theories.

\end{document}